\documentclass[traditabstract]{aa}  

\usepackage{graphicx}
\usepackage{txfonts}

\usepackage{epsfig}
\usepackage{natbib}

\def\lamb[#1]{#1\,{\AA}}
\def\lambr[#1-#2]{{{#1}--{#2}\,{\AA}}}
\def\rat[#1 #2]{#1/#2}
\def\serts89{SERTS-89}

\def\tabul{\hbox{\raise 0.75pt\hbox{$\triangleleft$}}}
\def\ergs[#1]{#1 {ergs}~{cm$^{-2}$}\,{s$^{-1}$}\,{sr$^{-1}$}}
\def\dens[#1]{10$^{#1}$\hskip 1.5pt{cm$^{-3}$}}
\def\densr[#1 #2]{10$^{#1}$\hskip 1pt{--}\hskip .5pt{10$^{#2}$}\hskip 1.5pt{cm$^{-3}$}}
\def\fl[#1 #2]{{#1}$\pm${#2}}
\def\orb[#1 #2]{{$#1^{#2}$}}
\def\ls[#1 #2]{{$^{#1}${#2}}}
\def\tm[#1 #2 #3]{{$^{#1}${#2}$_{#3}$}}

\begin{document}

\title{CHIANTI - An atomic database for Emission Lines. \\
 Version~8}
\author{G. Del Zanna\inst{1}
\and
{K.P. Dere}\inst{2}
\and
{P.R. Young}\inst{3}
\and
{E. Landi}\inst{4}
\and
{H.E. Mason}\inst{1}
}
\institute{DAMTP, Centre for Mathematical Sciences, University of Cambridge,
  Wilberforce road, Cambridge, CB3 0WA UK
\and
{School of Physics, Astronomy and Computational Sciences, MS 6A2, George Mason University, 4400 University Drive, Fairfax, VA 22030, USA}
\and
{College of Science, George Mason University, 4400 University Drive, Fairfax, VA, 22030} 
\and
{Department of Atmospheric, Oceanic and Space Sciences, University of Michigan, Ann Arbor, MI 48109}
}

  \date{Received  24/06/2015 ; accepted 12/08/2015}

  \abstract{We present version 8 of the CHIANTI database. 
This version includes a large amount of new 
data and ions, which represent a significant improvement in the soft X-ray, 
EUV and UV spectral regions,
which  several space missions  currently cover. 
New data for neutrals and low charge states are also added. 
The data are assessed, but to improve the modelling of low-temperature plasma
the effective collision strengths for most of the new datasets 
are not spline-fitted as previously, but are retained as calculated. 
This required a change of the  format of the CHIANTI electron excitation files.
The format of the energy files has also been changed.
Excitation rates between all the levels are 
 retained for most of the new datasets,
 so the data can in principle be used to model high-density plasma. 
In addition, the method for computing the differential emission measure 
used in the CHIANTI software has been changed.
\keywords{Atomic data -- Line: identification -- Atomic processes -- Radiation mechanisms: thermal }
}

\maketitle

\section{Introduction }


CHIANTI\footnote{www.chiantidatabase.org} is a database of assessed atomic parameters
 and transition rates needed  
for the calculation of the line and continuum  emission
 of optically thin, collisionally-dominated 
plasma. IDL-based software was initially developed to calculate synthetic spectra
 and measure  plasma  parameters such as electron densities and temperatures,
 and  is distributed within 
Solarsoft\footnote{www.lmsal.com/solarsoft/}.
A Python version of the CHIANTI software is also 
available\footnote{chiantipy.sourceforge.net/}.

CHIANTI was first released in 1996 \citep{dere_etal:97} 
 and since then various new releases  have been made available,
to  expand the database and improve the quality of the  data.
Particular emphasis has been given to the line identifications and 
improvement of the reference wavelengths. 
For several ions, the wavelengths and identifications are different from 
and more accurate than those of the NIST database \citep{nist:2013}, as recently 
confirmed by  laboratory measurements 
\citep{beiersdorfer_lepson:2012,beiersdorfer_etal:2014}. 
For the above reasons, 
 CHIANTI is now often used as a  reference atomic database for ions,
and is included in several other atomic codes and packages.

The version 8 of the database focuses on the inclusion of a large amount of
atomic data, much of which has been 
 produced by the UK APAP network\footnote{www.apap-network.org}. 
The new data are particularly relevant for current EUV/UV spectroscopic 
instruments aboard the 
Solar and Heliospheric Observatory (SOHO), Hinode, 
Solar Dynamics Observatory (SDO), and  
Interface Region
Imaging Spectrometer (IRIS) satellites.
They are also important for the X-rays (especially soft X-rays), observed by
Chandra and XMM-Newton. 
Version 8 also adds data for low charge states that were missing
and are relevant for modelling the ultraviolet (UV) spectral region.
  
The next Section describes the database improvements.
Section~3 describes the new CHIANTI DEM software  and 
shows as an example how the new data improves the analysis of SDO data.
 Section~4 provides the conclusions.

\section{Database improvements}

\subsection{New database format}

The formats of the core energy level (elvlc), radiative (wgfa) and
electron collision (splups) data files of CHIANTI have remained unchanged
since version 1 (Dere et al. 1997), except for a modification to the
splups files in version 4 \citep{young_etal:03} to allow 9-point spline fits
to the data in addition to the 5-point spline fits. The files are in ascii
format, with a fixed text width for each column and, in preparation for
the inclusion of atomic models with $\ge$ 1000 atomic levels, it has been
necessary to modify the format of the elvlc and splups files, which were
restricted to at most 999 levels. The wgfa files retain the same structure
as in previous versions of the database. The elvlc files have been
restructured (Appendix A.1) but retain the same name, while the splups
files have been significantly revised (Appendix A.2) and are now renamed
as "scups" files. For  both the updated elvlc and new scups files, an
ascii fixed text width format has been retained.

A fundamental change in philosophy for the electron collision data has
been implemented through the new scups files. Previously effective
collision strengths (either directly taken from the literature or
integrated from collision strengths) were scaled using the \cite{burgess_tully:92}
 method onto a scaled temperature $T_{\rm s}$ domain between 0 and
1, then either five or nine point spline were fit to the data in the
scaled domain. The five or nine point spline nodes were stored in the
splups files. For version 8 we continue to scale the effective collision
strength data, but now we directly store these scaled data points and no
longer perform the five or nine point spline fits. An extrapolation to
$T_{\rm s}=0$ is performed, and either the high temperature limit point
(Dere et al. 1997) or an extrapolation is used for the $T_{\rm s}=1$
value. This ensures that an effective collision strength can be derived
for any temperature between zero and infinity through interpolation of the
stored data points.

The advantage of the new format is that the stored CHIANTI data is now a
direct representation of the original data and there is no need to omit
data points as was sometimes necessary if the collision strength structure
could not be represented accurately by a nine point spline fit.
This often occurred at low temperatures. 
Providing the actual rates represents an improvement in accuracy for 
studies of very low temperature plasma, such as those of photoionised plasma.
We continue to assess the collision data using a mixture of automatic
checking procedures and by-eye inspection.

Much of the new atomic data added to version 8 has been processed in this
way. For data added in earlier versions of the database and retained for
version 8 we have simply re-formatted the previous file formats. For
example, if a transition was represented by a nine point spline fit, then
these nine spline points are stored in the new scups file.

A further advantage of the new method is that it is much more efficient to
process a data-set, allowing the complete set of collision strengths for
an ion to be added. This is in contrast to earlier versions of CHIANTI for
which often only transitions involving a metastable level were included.
This advance will allow the CHIANTI atomic models to be used in a
high-density regime as found in, for example, laboratory plasmas.
When calculating the level populations, the rates are
 obtained, as before,  by interpolation in the scaled domain.


\subsection{He-like ions: \ion{C}{v}, \ion{N}{vi}, \ion{O}{vii}, \ion{Ne}{ix},
\ion{Mg}{xi}, \ion{Al}{xii}, \ion{Si}{xiii}, \ion{S}{xv}, \ion{Ar}{xvii},
\ion{Ca}{xix}, \ion{Fe}{xxv}, \ion{Ni}{xxvii}, \ion{Zn}{xxix}}

For the He-like ions the main changes have been to insert the UK APAP network
radiation-damped  $R$-matrix effective collision strength calculations of 
\cite{whiteford_etal:01}  into the database for the ions listed. 
 Collision strengths are
provided between all levels of a 49-level fine structure model that includes all
levels up through 1s5$l$.  These new collision strengths replace those in the
7.1 version of the CHIANTI database \citep{landi:2013}.  
  In the cases of \ion{C}{v}, \ion{O}{vii},
\ion{Mg}{xi},  \ion{Al}{xii}, \ion{Si}{xiii}, \ion{Ni}{xxvii}, the previous collision
strengths were taken from the distorted-wave calculation of
\cite{zhang_sampson:1987} who provided collision strengths between the 1s$^2$,
1s2s, and 1s2p fine-structure levels.  Collision strengths to n=3-5 were taken
from the calculations of \cite{sgc:1983} which were performed with a hydrogenic
approximation.  
The new data therefore represent an improvement, especially for 
the forbidden transitions, which are increased  by the effect of 
the  resonances (although at high temperatures increases are only of the 
order of 10\%).

For  \ion{S}{xv} and \ion{Ca}{xix}, the previous version of
CHIANTI used the $R$-matrix calculations of \cite{kimura:2003} that provided
collision strengths between the 1s$^2$, 1s2$l$ and 1s3$l$ levels. Excitation to
the 1s4$l$ and 1s5$l$ were the hydrogenic collision strengths of
\cite{sgc:1983}. 
 For \ion{N}{vi}, the version 7.1 collision strengths were
taken from the combined $R$-matrix and FAC (Flexible Atomic Code, see 
\citealt{gu_fac:2008})  calculations of
\cite{aggarwal_keenan_heeter:2009}. 
These calculation are
 of similar accuracy to the APAP calculations but the latter have been
inserted into the current database for consistency.

 For \ion{Ar}{xvii} and
\ion{Fe}{xxv}, version 7.1 already included the rates of
\cite{whiteford_etal:01}.  \ion{Zn}{xxix} is new to the CHIANTI database.

The version 7.1 energy level files, e.g. \emph{fe\_25.elvlc} have been
updated.  The basic theoretical energies have been obtained with AUTOSTRUCTURE
\citep{badnell_AS:2011} calculations.  These are used to provide level energies
and to provide approximate wavelengths in cases where  energies based on observed
spectral lines are unavailable.  Many of the observed energies in previous
versions have  been taken from the NIST database \citep{nist:2013} and these have
been updated to the latest version (5.1) of the NIST database.  Highly accurate
calculations of the energies of the six  1s2s and 1s2p fine structure levels
provided by \cite{artemyev:2005} have been included.

The updated energies have been used to calculate the wavelengths in the radiative
data files, e.g. \emph{fe\_25.wgfa}. Weighted oscillator strengths and
A-values are obtained from an AUTOSTRUCTURE calculation.  A-values originating
from a number of the 1s2$l$ levels, varying from ion to ion, have been retained from
version 7.1.

The effect of cascades following recombination into He-like levels of
C V, N VI, O VII, Ne IX, Mg XI and Si XIII was modeled by \cite{porquet_dubau:2000},
 and these data were added to CHIANTI v.6 \citep{dere_etal:09_chianti_v6}.
\cite{porquet_dubau:2000} provided total rates into the
individual $1s2l$ levels for six temperatures covering 1.6~dex in log
temperature. One problem with this data-set is that the CHIANTI models
for these ions contain 49 levels and the cascading through levels
8--49 which would produce emission lines is not modeled. In particular
the lines due to dielectronic recombination that are important at
X-ray wavelengths are not modeled and so it was necessary to retain the
dielectronic files for these ions (see \citealt{dere_etal:01}). This has the
consequence that emission lines from the $1s2l$ levels (with indices
2--7 in the CHIANTI model) had additional contributions due to
double-counting of the recombination cascades. 

In addition, the temperature coverage
of the \cite{porquet_dubau:2000} recombination rates was small, and not
suitable for the wide range of uses of the CHIANTI database. For these
reasons, the Porquet \& Dubau recombination rates have been replaced
with the level-resolved radiative recombination rates of \cite{badnell_rr:2006} 
 into all 49 levels of the CHIANTI model. The level-resolved
dielectronic recombination rates contained in the dielectronic models of
the ions have been retained from the earlier versions of the
database. As the  \cite{badnell_rr:2006}  data are direct rates into individual
levels, then the new CHIANTI model does not include contributions from
levels with $n>5$  in contrast to \cite{porquet_dubau:2000} who included
cascading from all values of $n$. These additional contributions
however are small.

For the remaining He-like ions, the previous CHIANTI models used
radiative recombination rates from \cite{mewe:1985} computed using
the prescription of \cite{mewe_gronenschild:1981}.  These rates are
available for the $1s2l$ levels and selected other levels, and they
include the contributions from cascading. They have been
replaced with the direct recombination rates of \cite{badnell_rr:2006} into
all levels of the CHIANTI models. The level-resolved dielectronic
rates from the previous CHIANTI models that are stored in the
dielectronic files are retained.

To reduce the size of the data files containing the radiative 
rates, we have retained for all the new datasets 
only transitions with a branching ratio greater than 10$^{-5}$.

\subsection{Li-like ions}

\cite{liang_badnell:2011}
have performed, within the UK APAP network, scattering 
calculations for electron-impact excitation of all Li-like ions from Be$^+$ to Kr$^{33+}$
using the radiation- and Auger-damped intermediate-coupling frame transformation $R$-matrix approach.
The target included 204 close-coupling (CC) levels, with valence (up to $n=5$), 
and core-electron excitations (up to $n=4$).
During the assessment of these datasets, we  have uncovered 
 small inconsistencies in a few ions between the highest temperature values 
of the published rates and their high-temperature limits.
These were ultimately due to a mistaken repetition of the last few 
collision strengths. 
The data have been 
corrected and the effective collision strengths recalculated
(Liang, priv. comm., 2013).
The targets  represent a considerable improvement over previous ones,
which typically contained a much smaller number of levels, 
either 15 or 40.

\subsubsection{\ion{C}{iv}, \ion{N}{v}, \ion{O}{vi}, \ion{Ne}{viii}, \ion{Mg}{x}, \ion{Al}{xi}, \ion{Si}{xii},
\ion{S}{xiv}, \ion{Ar}{xvi}, \ion{Ca}{xviii}, \ion{Cr}{xxii}, \ion{Fe}{xxiv}, \ion{Ni}{xxvi}, \ion{Zn}{xxviii}}

The basic ion structure is determined from an AUTOSTRUCTURE calculation that  provides theoretical energy levels, oscillator strengths, A-values and autoionization rates.  However, the energies, wavelengths and autoionization rates are taken from the calculations of Safronova (as reported in \cite{kato_97} or privately communicated), where available.  
As with the abundant He-like ions, we have used the direct radiative recombination rates of \cite{badnell_rr:2006} to account for the effect of radiative recombination on the level populations.  We have used AUTOSTRUCTURE to calculate the bound levels up to n=8 for which the direct recombination rates are available.

The n=6,7,8 levels that have been added do not have any direct collisional excitation,
so they are only  populated through recombination.
This means that for normal collisionally-excited plasma the intensities of any lines produced 
by these levels are significantly underestimated,
and should therefore not be used for spectral analysis.

We have included only the collision strengths of \cite{liang_badnell:2011} 
 from the 2s and 2p levels. 
The data have been spline-fitted over a reduced number of temperatures.

A large number of the observed energy levels have been taken from \cite{nist_v5}.
However, we found that for several important ions the NIST energies were in need 
of improvement. Details  on each ion are provided as comments within the energy files.

For \ion{O}{vi}, several wavelengths have been taken from experimental
measurements and are show in  Table~1.  These wavelengths have also
been used to provide the 'observed' energies for their respective upper levels.

For \ion{Ne}{viii}, the wavelength of the 1\orb[s 2]2{\it s} \tm[2 S 1/2] --
1\orb[s 2]2{\it p} \tm[2 P 3/2] has been measured by \cite{peter_judge_99} and
\cite{dammasch_99_ne_8}.  These measurements are consistent with each other and
have been used to determine the energy of the 1\orb[s 2]2{\it p} \tm[2 P 3/2]
level and have been used as the 'observed' wavelength.

\cite{peter_judge_99} have also measured the wavelength of the \ion{Mg}{x}
1\orb[s 2]2{\it s} \tm[2 S 1/2] - 1\orb[s 2]2{\it p} \tm[2 P 1/2] line.  This
measurement has been used to derive the 'observed' energy of the 
1\orb[s 2]2{\it p} \tm[2 P 1/2] level and the 'observed' wavelength of that line in
CHIANTI.

For \ion{Fe}{xxiv}, observed energies for the ground level through 1\orb[s 2]5{\it d} have been taken from \cite{giulio_fe_24}.

\begin{table}[!htbp]
\begin{center}
\caption{Experimental wavelengths for \ion{O}{vi}}
\footnotesize
\begin{tabular}{llll}
\hline\hline\noalign{\smallskip}
$\lambda$ (\AA) & Lower & Upper & Ref. \\
\noalign{\smallskip}\hline\noalign{\smallskip}
23.017 & 1\orb[s 2]2{\it p} \tm[2 P 1/2,3/2] & 1{\it s}2\orb[s 2] \tm[2 S 1/2] & \cite{gu_2005_o} \\
22.374 & 1\orb[s 2]2{\it s} \tm[2 S 1/2] & 1{\it s}2{\it s}($^3$S)2{\it p} \tm[4 P 1/2]  & \cite{gu_2005_o}  \\
22.019 & 1\orb[s 2]2{\it s} \tm[2 S 1/2] & 1{\it s}2{\it s}($^1$S)2{\it p} \tm[2 P 1/2,3/2] &  \cite{schmidt_04} \\
21.845 & 1\orb[s 2]3{\it s} \tm[2 S 1/2]  & 1{\it s}2{\it s}($^1$S)3{\it p} \tm[2 P 1/2]  & \cite{gu_2005_o} \\
21.672 & 1\orb[s 2]3{\it s} \tm[2 S 1/2]  & 1{\it s}2{\it s}($^3$S)3{\it d} \tm[4 P 3/2]   & \cite{gu_2005_o}  \\
\noalign{\smallskip}\hline
\end{tabular}
\normalsize
\end{center}
\label{tab:o_6}
\end{table}

In the Version 7 of the CHIANTI database, the atomic models for these ions were assembled 
largely between 1997 and 2001.  The model for \ion{Fe}{xxiv} was assembled in 2005.  
The database for these ions reflect the atomic data that was available at the time.  
Energy levels were taken from Version 2 of the NIST database and from the series of 
NIST publications such as \cite{martin_1990_s} for the sulphur ions.  
A-values were taken from \cite{Martin_1993_li} or derived from the oscillator strengths 
provided in the collision strength calculations.  
Autoionization rates were taken from \cite{vainstein_safronova_1978} that were updated by U. Safronova in 1999. 
Collision strengths were taken from a variety of sources.
For a few ions,  distorted-wave collision strengths calculated 
 by \cite{zhang_sampson_fontes_1990} for the bound levels up to 5g were included,
together with  inner-shell excitation rates provided by \cite{goett_sampson_1983}. 
\ion{C}{iv} and \ion{N}{v} only included 15 levels, up to 4f.
 The data provided in Version 8 therefore  marks a considerable improvement over the previous Version.

To reduce the size of the data files containing the radiative 
rates, we have retained for all the new datasets 
only transitions with a branching ratio greater than 10$^{-5}$.

\subsubsection{\ion{Na}{ix}, \ion{P}{xiii}, \ion{Cl}{xv}, \ion{K}{xvii}, \ion{Mn}{xxiii}, and \ion{Co}{xxv}}

For these ions we have retained the original 204 levels and the original 
energies, A-values and all the effective collision strengths as calculated by \cite{liang_badnell:2011}.
To reduce the size of the data files containing the radiative 
rates, we have retained for all the new datasets 
only transitions with an A-value greater than 10$^{-5}$ the 
largest A-value (for each level).

We have calculated the autoionization rates using AUTOSTRUCTURE 
(version 24.24)  and 
the same set of scaling parameters used by \cite{liang_badnell:2011} 
for consistency.
For \ion{Na}{ix}, experimental energies have been reassessed on the basis of the 
wavelength measurements as reported in \cite{kelly:87}.
For \ion{P}{xiii}, \ion{Cl}{xv}, \ion{K}{xvii}, \ion{Mn}{xxiii}, and \ion{Co}{xxv}
experimental energies have been  taken from \cite{nist_v5}.

In the Version 7 of the CHIANTI database, 
 \ion{Na}{ix},\ion{P}{xiii}, \ion{K}{xvii}, \ion{Mn}{xxiii}, and \ion{Co}{xxv}
included only 20 levels up to 5d, and 
DW collision strengths were provided by \cite{zhang_sampson_fontes_1990}
\ion{Cl}{xv} is a new entry in the database.


\subsection{B-like ions}

\cite{liang_etal:2012} have performed, within the APAP network,
ICFT  $R$-matrix calculations 
for electron-impact excitation amongst  204 close-coupling levels 
 for all boron-like ions from C$^+$ to Kr$^{31+}$.
We include the \cite{liang_etal:2012} collision strengths and A-values
for several ions, as described below.
We note, however, that we have replaced the A-values of the transitions
among the lowest levels as calculated by \cite{liang_etal:2012}
with  {\sc autostructure} with more accurate values, whenever available. 
For example, those calculated by \cite{correge_hibbert:2004} with the CIV3
code \citep{hibbert:1975} and those calculated by \cite{rynkun_etal_2012} with 
the multi-configuration Dirac-–Hartree-–Fock (MCDHF)  GRASP2K code \citep{grasp2k}.
This is because the ab-initio energies of the CIV3 and GRASP2K are normally
more accurate than those by  {\sc autostructure}, and so are the A-values.

The \ion{Fe}{xxii} ion model included the \cite{badnell_etal:01} 
collision strengths so it is left unchanged, since the 
 \cite{liang_etal:2012} calculation is essentially the same.

During the present assessment, we have uncovered an error in the data.
The 2s 2p$^2$ $^2$S$_{1/2}$, $^2$P$_{1/2}$
levels (No. 8,9) were inverted by mistake, hence the 
collision strengths and A-values for transitions connected to these levels were incorrect.
The data for these transitions  have been recalculated
(Liang, priv. comm., 2014).
To reduce the size of the data files containing the radiative 
rates, we have retained for all the new datasets 
only transitions with an A-value greater than 10$^{-5}$ the 
largest A-value (for each level).

\subsubsection{\ion{C}{ii}}

\ion{C}{ii} produces  a number of strong emission lines in the UV, including the 
2s$^2$ 2p $^2$P$_J$ - 2s 2p$^2$ $^2$D$_{J'}$ multiplet 
found between 1334.5 and 1335.7~\AA\  that is
routinely observed by the recently-launched IRIS instrument 
\citep{depontieu_etal:2014}.
For \ion{C}{ii}, the  collisional data of \cite{liang_etal:2012}
replace those of  \cite{tayal:2008}, which included only 37 levels.
Observed energies are taken from NIST \citep{nist_v5}.
A-values are taken from \cite{liang_etal:2012}, with the exception of 
several transitions of the lower levels, where the 
\cite{correge_hibbert:2004} data are used.
We note that relatively good agreement is found between these 
A-values and those of  \cite{liang_etal:2012},  but not with those of 
\cite{tayal:2008}.


\subsubsection{\ion{N}{iii}}

For \ion{N}{iii}, the  \cite{liang_etal:2012}
collisional data 
replace those of  \cite{stafford_etal:1994} which included only 20 levels.
This is a significant improvement. 
A-values are taken from \cite{liang_etal:2012}, 
with the exception of the transitions of the lower 20 levels
(2s 2p$^2$, 2p$^3$ , 2s$^2$ 3l)
calculated by \cite{correge_hibbert:2004}, which were already 
present in the previous version of CHIANTI.
Observed energies have been taken from NIST \citep{nist_v5}.

\subsubsection{\ion{O}{iv}}

\ion{O}{iv} is an important ion for solar physics applications,
in particular for measuring electron densities.
The  \cite{liang_etal:2012}
collisional data  for \ion{O}{iv} replace those of 
\cite{aggarwal_keenan:2008} which included only 75 levels.

There is overall agreement between the two calculations, with 
differences of the order of 10\%  for the 
important transitions in the IRIS wavelength range 
(see \citealt{dudik_etal:2014_o_4}).


Observed energies have been taken from the previous 
CHIANTI version and from NIST \citep{nist_v5}, except the 2s 2p$^2$,
which have been revised.
A-values are taken from \cite{liang_etal:2012}, 
with the exception of the transitions among the lower 20 levels
(2s 2p$^2$, 2p$^3$ , 2s$^2$ 3l)
calculated by \cite{correge_hibbert:2004}, which were already 
present in the previous version of CHIANTI.

\subsubsection{\ion{Ne}{vi}}

The  \cite{liang_etal:2012}
collisional data  replace those from a similar calculation, an 
180-level intermediate-coupling frame-transformation (ICFT)
 $R$-matrix close-coupling calculation by 
\cite{mitnik_etal:2001}. 
Observed energies have been taken from NIST \citep{nist_v5}.
A-values are taken from \cite{liang_etal:2012}
with the exception of the transitions among the lower 15 levels
(2s 2p$^2$, 2p$^3$) for which we have adopted the results of the 
MCDHF  calculations obtained by  
\cite{rynkun_etal_2012} with the GRASP2K code.


\subsubsection{\ion{Mg}{viii}}

For \ion{Mg}{viii}, we have adopted the  
$R$-matrix ICFT  collisional data of \cite{liang_etal:2012}.
The new collisional data replace the DW data of 
\cite{zhang_sampson:1994a}, which only contained excitation 
within the 15 $n=2$ levels, and unpublished calculations 
from Sampson and Zhang (1995) for the $n=3$ levels.
Observed energies have been taken from NIST \citep{nist_v5}.
A-values are taken from \cite{liang_etal:2012}
with the exception of the transitions among the lower 15 levels
(2s 2p$^2$, 2p$^3$) for which we have adopted the results of the 
MCDHF calculations obtained by  
\cite{rynkun_etal_2012} with the GRASP2K code.

\subsubsection{\ion{Si}{x}}

\ion{Si}{x} is a particularly important ion for measuring 
electron densities of the solar corona (see e.g. an example in 
\citealt{delzanna:12_atlas}).
The \cite{liang_etal:2012} data replace the 
\cite{liang_etal:2009_si_10} ICFT $R$-matrix calculation, which  included 125 
fine-structure levels (58 LS terms) belonging to the spectroscopically-important 
$n=2,3$ configurations. 
The experimental energies for this ion were assessed for version
7 \citep{landi_v7}, and have been taken from a variety of sources. 
A-values are taken from \cite{liang_etal:2012}
with the exception of the transitions among the lower 15 levels
(2s 2p$^2$, 2p$^3$) for which we have adopted the results of the 
MCDHF calculations obtained by  
\cite{rynkun_etal_2012} with the GRASP2K code.

\subsubsection{ \ion{Al}{ix},  \ion{S}{xii}, \ion{Ar}{xiv}, and \ion{Ca}{xvi}}

\ion{Al}{ix},  \ion{S}{xii},  \ion{Ar}{xiv}, and \ion{Ca}{xvi}
are ions that are particularly important for solar physics diagnostics.
For example, \ion{S}{xii}, \ion{Ar}{xiv}, and \ion{Ca}{xvi} produce lines in the EUV
 observed with the  Hinode EUV Imaging Spectrograph (EIS),  and that 
have recently been used for elemental abundance studies 
(cf. \citealt{delzanna:2013_multithermal}).

The ion models for 
\ion{Al}{ix},  \ion{S}{xii},  \ion{Ar}{xiv}, and \ion{Ca}{xvi}, 
as described in version 1 \citep{dere_etal:97}, contained several datasets,
with some unpublished data. In terms of collision strengths, 
they included  $R$-matrix calculations of \cite{zhang_etal:1994} for the 
15 fine-structure levels arising from the $n=2$ configurations.
Collision strengths to the 110 fine-structure $n=3$ levels were  
unpublished,  taken from a 
simple Coulomb-Born calculation by Zhang and Sampson (1995). 
An error in the $R$-matrix calculations of \cite{zhang_etal:1994}
affecting  \ion{Al}{ix},  \ion{Si}{x}, \ion{S}{xii} \ion{Ar}{xiv}, 
and \ion{Ca}{xvi}  was uncovered, and corrected rates were introduced in 
version 4 \citep{young_etal:03}.

We have replaced the previous collision rates with the 
\cite{liang_etal:2012} APAP data.
As noted by  \cite{liang_etal:2012}, the new data 
have significantly larger collision strengths for  the $n=2$ levels, due to the extra 
resonances attached to the $n=3$ levels, and which were not included in the 
\cite{zhang_etal:1994} calculations. 
 As shown in  Fig.4 of \cite{liang_etal:2012}, 
for  \ion{Ar}{xiv} there is good agreement (within a relative 20\%) only
 for the strongest transitions; however, the effect of the resonances increases
the collision strengths by about a factor of two or more 
for the weaker transitions.

Observed energies have been taken mostly from NIST \citep{nist_v5},
 although we note that 
in several cases values are only known relative to the energies of other levels. 
A-values are taken from \cite{liang_etal:2012}
with the exception of the transitions among the lower 15 levels
(2s 2p$^2$, 2p$^3$) for which we have adopted the results of the 
MCDHF calculations obtained by  
\cite{rynkun_etal_2012} with the GRASP2K code.

For \ion{Al}{ix}, \ion{Ar}{xiv}, and \ion{Ca}{xvi}
 we used the corrected collisional data involving the two levels (8,9).

For \ion{S}{xii}, the intensities of the strongest EUV lines are very similar to those
of the previous model.
The energies of some \ion{S}{xii} levels are taken from \cite{lepson_etal:2005}.




\subsubsection{ \ion{Na}{vii},, \ion{P}{xi}, \ion{K}{xv}, \ion{Cr}{xx},  \ion{Mn}{xxi},  \ion{Co}{xxiii}}

For  \ion{Na}{vii}, \ion{P}{xi}, \ion{K}{xv}, \ion{Cr}{xx},  \ion{Mn}{xxi}
and  \ion{Co}{xxiii}
 the \cite{liang_etal:2012} collisional data replace the DW data of 
\cite{zhang_sampson:1994a}, which only contained excitation 
within the 15 $n=2$ levels.
Observed energies have been taken from NIST \citep{nist_v5}, although we note that 
in several cases values are only known relative to the energies of other levels. 

For \ion{Na}{vii},  \ion{P}{xi}, \ion{K}{xv} 
we used the corrected collisional data involving the two levels (8,9).

Observed energies have been taken from NIST \citep{nist_v5}.
A-values are taken from \cite{liang_etal:2012}
with the exception of the transitions among the lower 15 levels
(2s 2p$^2$, 2p$^3$) for which we have adopted the results of the 
MCDHF calculations obtained by  
\cite{rynkun_etal_2012} with the GRASP2K code.




 \subsubsection{\ion{Cl}{xiii},  \ion{Ni}{xxiv}}

\ion{Cl}{xiii} is  a new entry to the database.
We used the collision rates of \cite{liang_etal:2012},
with the corrected data involving the two levels (8,9).
Observed energies have been taken  from NIST \citep{nist_v5}.
A-values are taken from \cite{liang_etal:2012}
with the exception of the transitions among the lower 15 levels
(2s 2p$^2$, 2p$^3$) for which we have adopted the results of the 
MCDHF calculations obtained by  
\cite{rynkun_etal_2012} with the GRASP2K code.

For \ion{Ni}{xxiv},  the \cite{liang_etal:2012} collisional data replace 
those  calculated by \cite{sampson_etal:1986}.
Observed energies have been taken from NIST \citep{nist_v5}.
A-values are taken from \cite{liang_etal:2012}
with the exception of the transitions among the lower 15 levels
(2s 2p$^2$, 2p$^3$) for which we have adopted the results of the 
MCDHF calculations obtained by  
\cite{rynkun_etal_2012} with the GRASP2K code.

 \subsection{C-like ions}

\subsubsection{\ion{C}{i}}

\ion{C}{i} is a new addition to the database. 
\cite{wang_etal:2013} have recently performed a 
$B$-spline $R$-matrix calculation for electron-impact excitation and ionization of carbon.
\cite{wang_etal:2013} provided $LS$-resolved effective collision strengths,
but only covering high temperatures (above 10,000 K). 
One of the authors (O. Zatsarinny) has recalculated the 
$LSJ$-resolved effective collision strengths   over a more extended
temperature range (from 1000 to 500 000 K), and also provided us with the 
level-resolved oscillator strengths and 
A-values for the E1, E2 and M2 transitions, which we have included. 
We have verified that the effective collision strengths converge 
to the high-energy limits for the allowed transitions.

A-values  for the forbidden lines within the ground configuration 
are taken from the CIV3 calculations of \cite{hibbert_etal:1993}.
Experimental energies have been taken from \cite{chang_geller:1998}.

\subsection{N-like ions}

\subsubsection{\ion{N}{i}}

We have included new effective collision strengths and gf-values
for \ion{N}{i} from \cite{tayal:2006}.
The collision strengths were obtained with the $R$-matrix B-spline method, 
 including 24 spectroscopic bound and autoionising states together with 15 pseudostates. 
This calculation is much larger than the 18-states previous 
one  from \cite{tayal:2000}, which was 
added in CHIANTI version 4 \citep{young_etal:03}.

A-values for the allowed and forbidden lines have been taken from \cite{tachiev_etal:2002}.
They were calculated with the energy-adjusted  MCHF  method.
Observed energies have been taken from NIST \citep{nist_v5}.


\subsection{O-like ions}


\subsubsection{\ion{Fe}{xix}}

For \ion{Fe}{xix} we have included the collision strengths from the 
$R$-matrix calculations of \cite{butler_badnell:2008}. 
The target included 342 close-coupling levels up to $n=4$.
They replace those calculated with the DW approximation and the FAC code  by 
\cite{gu:2003}. 
We have retained the 
extra levels (from 343 to 636) of the previous CHIANTI model, which included 
data from \cite{landi_gu:2006} and collisional data (above ionization) from 
\cite{bautista_etal:2004}.
We have also merged the A-values. Those for the allowed transitions 
of the lowest  342 close-coupling levels are from  \cite{butler_badnell:2008},
while the remainder  are from the previous CHIANTI model.

\subsection{F-like ions}

\cite{witthoeft_etal:07_f-like} calculated, within the APAP network, 
ICFT  $R$-matrix calculations 
for electron-impact excitation amongst 195 $n=3$ close-coupling levels 
 for all F-like ions.
 The data for \ion{Si}{vi}, \ion{S}{viii}, \ion{Ar}{x}, \ion{Ca}{xii},
\ion{Fe}{xviii}, and  \ion{Ni}{xx} 
were already added in  CHIANTI version 6 
\citep{dere_etal:09_chianti_v6}.

For version 8 we have added the data for 
  \ion{Na}{iii},  \ion{Mg}{iv}, and \ion{Al}{v}. 
The data for these lower charge state ions were not originally included
in CHIANTI because they are of more limited accuracy than the data 
for the other ions in the sequence. However, these data are still 
a significant improvement,
considering that previous CHIANTI versions only had three 
levels, and the excitation data for \ion{Na}{iii} and \ion{Al}{v} 
were interpolated along the sequence.

For \ion{Ne}{ii}, we have included the collision strengths 
calculated by \cite{griffin_etal:2001}. As noted by 
\cite{witthoeft_etal:07_f-like}, the \cite{griffin_etal:2001} 
close-coupling expansion was smaller (138 levels), however the 
target was more accurate because pseudo-orbitals were used to 
improve the level energies.
The A-values for \ion{Ne}{ii} for the dipole-allowed transitions
have been taken from \cite{griffin_etal:2001}, while 
A-values for a few M1, M2 , E2 forbidden transitions
have been taken from an ab-initio MCHF calculation by 
G. Tachiev and C. Froese Fischer (2002), see 
the MCHF collection\footnote{http://nlte.nist.gov/MCHF/} at NIST.

To reduce the size of the data files containing the radiative 
rates, we have retained for all the new datasets 
only transitions with an A-value greater than 10$^{-5}$ the 
largest A-value (for each level).
Observed energies have been taken from NIST \citep{nist_v5}.

%
%


\subsection{Ne-like ions}

\cite{liang_badnell:2010}
calculated electron-impact excitation of all Ne-like ions from Na$^+$ to Kr$^{26+}$ 
using the ICFT $R$-matrix approach, for a large target
comprising of 209 levels, up to outer-shell promotions to $n=7$.
These calculations represent a significant improvement over the few 
previous calculations along the sequence.
In general, we have adopted the  \cite{liang_badnell:2010} energies, 
A-values and collision strengths. 
To reduce the size of the data files containing the radiative 
rates, we have retained for all the new datasets 
only transitions with an A-value greater than 10$^{-5}$ the 
largest A-value (for each level).
The experimental energies are from NIST \citep{nist_v5},
with few exceptions, most of which are noted below.

We have added the following datasets, which represent new ions 
not previously  present in the CHIANTI database:
 \ion{Na}{ii},  \ion{Mg}{iii}, \ion{Al}{iv}, \ion{P}{vi}, \ion{K}{x}, \ion{Ti}{xiii},
\ion{Cr}{xv},  \ion{Mn}{xvi},  \ion{Co}{xviii} and \ion{Zn}{xxi}.

We have also replaced the previous model ions for  \ion{Si}{v} and \ion{S}{vii}.
The previous \ion{Si}{v} and \ion{S}{vii}  model ions only had 27 levels, 
the ground configuration 2p$^6$ and the 2p$^5$ 3s, 2p$^5$ 3p, 2p$^5$ 3d
levels. 
In the case of  \ion{Si}{v}, the 
 previous collision strengths were calculated with the
 DW approximation.
In the case of \ion{S}{vii}, we have adopted for most
of the energies the experimental values of \cite{jupen_engstrom:2002},
while some are  from NIST \citep{nist_v5}.

Finally, we have replaced the previous model ions for \ion{Ar}{ix},  \ion{Ca}{xi}, and \ion{Ni}{xix}.
The previous CHIANTI models for these ions 
had collision strengths from  \cite{zhang_etal:1987},
calculated using a Coulomb-Born-Exchange method for the 
88 levels arising from $n=3, 4$ configurations.

For \ion{Ca}{xi}, we have assessed the \cite{ragozin_etal:1988}
experimental energies, but found various inconsistencies, so we have adopted
the  NIST \citep{nist_v5} values instead.
The  important \ion{Fe}{xvii} data were released previously 
in CHIANTI version 7 \citep{landi_v7}.

As in the case of the Li-like ions mentioned above, some of the collision strengths 
calculated by \cite{liang_badnell:2010} 
were affected by a  minor error which only affected  the high-temperature rates. 
We used the corrected data (as obtained from the authors) 
for \ion{Mg}{iii}, \ion{P}{vi}, \ion{Mn}{xvi}, \ion{Si}{v},  \ion{S}{vii},
\ion{Ar}{ix}, and \ion{Ni}{xix}.

For the astrophysically important ions 
\ion{Mg}{iii}, \ion{Al}{iv}, \ion{Si}{v},   
 \ion{S}{vii}, \ion{Ar}{ix}, \ion{Ca}{xi}, and \ion{Ni}{xix}
we have replaced the \cite{liang_badnell:2010} A-values of 
the transitions among the  2p$^{6}$ and 2p$^{5}$3l configurations 
(lowest 27 levels) with the  results of the 
MCDHF calculations obtained by    \cite{jonsson_etal:2014}  with the GRASP2K code
\citep{grasp2k}.
For some ions, the ordering of the levels was different,
as well as the LSJ designation of the dominant term.
For all the strongest transitions, we found excellent
agreement (to within a few percent) between the 
\cite{liang_badnell:2010} and \cite{jonsson_etal:2014} A-values.


\subsection{Na-like ions}

Within the UK APAP network, \cite{liang_etal:2009} calculated 
outer-shell electron-impact excitation of all Na-like ions from Mg$^+$ to Kr$^{25+}$ 
using the ICFT $R$-matrix approach.
The target and close-coupling expansion included 18 $LS$ terms of configurations
up to $n=6$. 
\cite{liang_etal:2009b} presented the inner-shell 
 electron-impact excitation of all Na-like ions from Mg$^+$ to Kr$^{25+}$
with the ICFT $R$-matrix  method with both Auger and radiation damping 
included via the optical potential approach. 
We have included these datasets for the astrophysically-important ions. The radiative data 
are taken from the same sets of calculations.
To reduce the size of the data files containing the radiative 
rates, we have retained for all the new datasets 
only transitions with an A-value greater than 10$^{-5}$ the 
largest A-value (for each level).

We have calculated (using {\sc autostructure} version 24.24) the autoionization rates 
for the levels above ionization, using the same set of configurations
and the same scaling parameters as adopted by \cite{liang_etal:2009b}.

For \ion{Al}{iii}, \ion{Si}{iv}, \ion{P}{v}, \ion{S}{vi}, \ion{Ar}{viii}, \ion{K}{ix},
 \ion{Ca}{x}, \ion{Cr}{xiv},  \ion{Mn}{xv}, \ion{Fe}{xvi}, \ion{Co}{xvii}, \ion{Ni}{xviii}
the new $R$-matrix results replace the 
DW collision strengths of \cite{sampson_etal:1990}, and the inner-shell
transitions  are a new addition.

For \ion{Mg}{ii}, the new data replace the  $R$-matrix 
calculations of \cite{sigut_pradhan:1995}, which included 10 $LS$ 
target terms. The method of calculation for  this older calculation is very similar
to the \cite{liang_etal:2009}, so the latter results should be more accurate,
given the larger target employed.
We note that the \ion{Mg}{ii}  h and k lines have a particular relevance for the 
recent solar spectroscopic mission IRIS; however, their emissivities
are only slightly changed (of the order of 10\%).

For \ion{Ti}{xii} and \ion{Zn}{xx}, the   \cite{liang_etal:2009} data 
replace the DW collision strengths  calculated by \cite{sampson_etal:1990}.



\subsection{Si-like ions}

\subsubsection{\ion{S}{iii}} %

New collisional data for \ion{S}{iii} have been calculated with the 
$R$-matrix suite of codes  by 
\cite{hudson_etal:2012}, and have been included.
They replace the previous $R$-matrix  calculations from 
\cite{tayal_gupta:1999}.
We have retained the A-values for the allowed and forbidden 
transitions we had in the previous CHIANTI version, 
calculated by \cite{froese_fischer:2006} 
with the ab-initio MHCF method, and the \cite{tayal_gupta:1999} 
A-values for the decays from the 4d levels.
Observed energies have been taken from NIST \citep{nist_v5}.

We have compared the intensities of the brightest lines 
with the new model ion, and found some small increases 
(of the order of 20\%) for some transitions, due to the extra
resonances related to the larger target.

\subsubsection{\ion{Fe}{xiii}} 

The previous \ion{Fe}{xiii} model included  electron impact excitation rates 
calculated by  \cite{storey_zeippen:2010} with the $R$-matrix method
 for a total of 114 fine-structure levels,
as well as DW data for the higher levels, up to $n=5$.
 
The new model includes APAP data from a
much larger  $R$-matrix calculation
which included a total of 749 levels up to $n=4$ \citep{delzanna_storey:2012_fe_13}.
As discussed in \cite{delzanna_storey:2012_fe_13}, cascading from 
 $n=5,6$ levels has a  small effect for the strongest lines, so it is left
out in the present model, to reduce the size of the ion.

The intensities of the transitions from the $n=3$ levels (in the EUV) are in close
agreement with those of the previous model. The new model provides, however,
 more accurate data 
for the soft X-ray lines (decays from the  $n=4$ levels) as compared to the 
previous version 7.1, which had collision strengths calculated with the 
DW approximation (see \citealt{delzanna_storey:2012_fe_13} for details).
Observed energies have been taken from  the benchmark works 
of \cite{delzanna:11_fe_13} and \cite{delzanna:12_sxr1},
where several new identifications in the EUV and in the soft X-rays were provided.
Empirically-adjusted theoretical wavelengths as obtained 
by \cite{delzanna_storey:2012_fe_13} are adopted.

\subsubsection{\ion{Ni}{xv}}

 \ion{Ni}{xv} produces strong decays from  the 3s 3p$^3$ and
3s$^2$ 3p 3d configurations
in the EUV region of the spectrum, several of which are observed by Hinode EIS,
and are useful for density and abundance diagnostics of solar active regions
\citep{delzanna:2013_multithermal}.
It also produces   soft X-ray  transitions  from the $n=4$ levels.

We have included the recent APAP $R$-matrix calculations 
from \cite{delzanna_etal_2014_ni_15}. They were obtained with 
a large target (up to  $n=4$ levels), similar to the one 
adopted for  \ion{Fe}{xiii}.
The new data replace the DW calculations of \cite{landi_bhatia:2012}.

\subsection{P-like ions}

\subsubsection{\ion{S}{ii}}

We have replaced the previous model with new collision strengths from 
\cite{tayal_zatsarinny:2010}. 
The A-values for the E1 transitions are taken from the paper,
while the A-values of the E2 and M1 transitions have been 
calculated by  O. Zatsarinny (priv. comm.).
Observed energies have been taken from NIST \citep{nist_v5}.

\subsubsection{\ion{Fe}{xii}} 

 The previous \ion{Fe}{xii} model included  electron impact excitation rates 
calculated by \cite{storey_etal:04} with the $R$-matrix codes 
for the lowest 143 levels, and DW data for the higher levels up to n=5.

The new model includes APAP data from a much larger $R$-matrix calculation
which included 912 levels up to  $n=4$
 \citep{delzanna_etal:12_fe_12}. 
The combined effect of extra cascading 
and increased excitation resulted in significantly (and surprisingly) 
different level populations of the levels of the ground configuration, 
which affected not only the forbidden lines, 
but also  the electron density measurements obtained with 
Hinode EIS. The new model provides densities about a factor of three lower
then the  \cite{storey_etal:04} model.
The intensities of the decays from the 3s 3p$^4$, observed in the 
EUV, are also  increased.
The extra cascading due to the $n=5,6$ levels was not found to be significant
for the strongest transitions
(less than 10\%, see \citep{delzanna_etal:12_fe_12})
 so is not included in the present model,
to reduce the size of the ion.

Observed energies have been taken from \cite{delzanna_mason:05_fe_12}
and from \cite{delzanna:12_sxr1} for the soft X-ray lines.
where several new identifications were provided.
Empirically-adjusted theoretical wavelengths as obtained 
by  \cite{delzanna_etal:12_fe_12} are used.

\subsection{S-like ions}

 \subsubsection{\ion{S}{i}}

\ion{S}{i} is a new addition to the database. 
We have included effective collision strengths from \cite{tayal:2004}.
gf and E1  A-values have been obtained from \cite{zatsarinny_bartschat:2006}. 
We have noticed significant discrepancies between the 
two datasets for the allowed transitions.
We have included the A-values for the forbidden transitions as 
calculated 	with  the ab-initio MHCF method 
by Irimia and Froese Fischer (unpublished, 2003,
but as available at the 
MHCF database\footnote{http://nlte.nist.gov/MCHF/view.html}).
Experimental energies have been taken from NIST \citep{nist_v5}.

\subsubsection{\ion{Fe}{xi}}

The previous \ion{Fe}{xi} model included  APAP electron impact excitation rates 
calculated by  \cite{delzanna_etal:10_fe_11} with the $R$-matrix codes 
for the lowest 365 levels, and DW data for  higher levels up to $n=5$.
The target adopted by \cite{delzanna_etal:10_fe_11} 
was an ad-hoc one, chosen to provide accurate  collision strengths for
transitions to three $n$=3, $J$=1  levels, which produce
some of the strongest EUV lines for this ion. 
These levels have a  strong spin-orbit interaction, and the 
collision strengths to these levels are very sensitive to the target. 
The accuracy of the target was confirmed a posteriori with  comparisons against observations
\citep{delzanna:10_fe_11}.

The new model includes APAP data from a much larger $R$-matrix calculation,
which included 996  levels up to $n=4$  \citep{delzanna_storey:2013_fe_11}.
This calculation significantly improved the atomic data for many lower 
(within the $n$=3) and higher 
($n$=4) levels.
As in the  \ion{Fe}{xii} case, \cite{delzanna_storey:2013_fe_11} found 
that a number of lower $n=3$ levels had  enhanced populations, 
caused by increased cascading from higher  levels 
whose collision strengths are enhanced compared to the previous calculation.

The larger target, however, 
did not produce accurate collision strengths for the three $n$=3, $J$=1  levels,
for which  we have retained the 
collision strengths calculated by  \cite{delzanna_etal:10_fe_11}.

The new larger calculations significantly affected the predicted intensities of several
strong EUV lines, some of which are  observed by Hinode EIS.
Experimental energies have been taken from 
\cite{delzanna:10_fe_11} and from  \cite{delzanna:12_sxr1},
where several new identifications  for the soft X-ray lines were provided.
Empirically-adjusted theoretical wavelengths as obtained 
by  \cite{delzanna_storey:2013_fe_11} are adopted.

\subsection{Cl-like ions}

\subsubsection{\ion{Fe}{x}}

The new model ion includes a large-scale
 APAP $R$-matrix calculation \citep{delzanna_etal:12_fe_10}
including 552 levels up to $n=4$. It replaces a previous 
 APAP $R$-matrix calculation \citep{delzanna_etal:04_fe_10}
which included 54 levels up to $n=3$, and a DW calculation
up to  $n=5$.

The new atomic calculations for the \ion{Fe}{x} 3s$^2$ 3p$^4$ 4s levels 
showed that DW calculations significantly underestimate 
the collision strengths,  as discussed in \cite{delzanna_etal:12_fe_10}.

The new larger calculations also significantly affect the predicted intensities of some
EUV lines, in particular the strong  decays from the 
3s$^2$ 3p$^4$  3d  configuration observed by Hinode EIS (the 257.26~\AA\ self-blend), 
and from  the  3s 3p$^6$ $^2$S$_{1/2}^{\rm e}$ level.
Also, the intensities of the   visible forbidden lines are significantly 
changed,  as shown in \cite{delzanna_etal:2014_fe_9}.

Experimental energies have been taken from \cite{delzanna_etal:04_fe_10}
and from \cite{delzanna:12_sxr1} for the soft X-ray lines.
Empirically-adjusted theoretical wavelengths as obtained 
by \cite{delzanna_etal:12_fe_10} are adopted.

\subsection{Ar-like ions}

\subsubsection{\ion{Fe}{ix}}

We have included the recent  APAP $R$-matrix calculations from 
\cite{delzanna_etal:2014_fe_9}, which replace the earlier 
calculations by \cite{storey_etal:02}. The new data were obtained with a large 
target, which included the main levels up to $n=5$.
\ion{Fe}{ix} produces strong lines in the 
Hinode EIS spectral range,  from the 3s$^2$ 3p$^4$ 3d$^2$ and
3s$^2$ 3p$^5$ 4p configurations. They  can be used to measure electron 
temperatures \citep{young:09, delzanna_etal:2014_fe_9}.
The intensities of these lines are only slightly different, with the 
exception of the 197.854~\AA\ line, for which significant changes are present,
as discussed in  \cite{delzanna_etal:2014_fe_9}.
The new calculations also improve the data for several 
soft X-ray lines, decays from the $n=4,5$ levels, for which 
the previous CHIANTI version had DW data from \cite{odwyer_etal:12_fe_9}.
We have retained the collision strengths and A-values of 
\cite{odwyer_etal:12_fe_9} for the  3s$^2$ 3p$^5$ 6$l$ ($l$=s,p,d,f,g)
configurations, by merging the two datasets.

\subsubsection{\ion{Ni}{xi}}

We have included the recent  APAP $R$-matrix calculations from 
\cite{delzanna_etal_2014_ni_11}. 
The target  included the main configurations up to $n=4$, and is a significant
improvement over the earlier $R$-matrix (only three  $n=3$ configurations,
calculated by \citealt{aggarwal_keenan:2007}) 
and DW calculations \citep{bhatia_landi:2011} which were included in the 
previous CHIANTI version. 
The two \ion{Ni}{xi} lines at 207.9 and
 211.4~\AA\ are observed by Hinode EIS. 
They are analogues of the \ion{Fe}{ix} 241.7, 244.9~\AA\ lines,
 which are an excellent density diagnostic.  
The experimental energies discussed by \cite{delzanna_etal_2014_ni_11}
are adopted.

\subsection{K-like ions}

\subsubsection{\ion{Fe}{viii}}

The collision strengths for \ion{Fe}{viii} 
have been notoriously difficult to calculate accurately.
The previous version if the CHIANTI database contained the 
\cite{griffin_etal:00} electron excitation rates, scaled
according to the benchmark structure calculations 
described in \cite{delzanna:09_fe_8}. The adjustment 
improved agreement with observations of the EUV lines but was clearly not 
satisfactory. A new scattering calculation was carried out 
by \cite{tayal:11_fe_8} and significantly improved the \cite{griffin_etal:00}
data.  However, as pointed out in 
\cite{delzanna_badnell:2014}, the \cite{tayal:11_fe_8} calculation did not improve
the rates for all the transitions, in particular for some
currently observed by the Hinode EIS spectrometer. 
A new large-scale (up to $n=5$, 518 levels) $R$-matrix ICFT calculation 
 has recently been carried out by \cite{delzanna_badnell:2014},
adopting a new method which employs semi-empirical corrections based on the 
term energy corrections. 
The new data were shown by \cite{delzanna_badnell:2014}
 to substantially improve agreement for the EUV lines,
some of which are useful  to measure electron temperatures
\citep{delzanna:09_fe_8}.
The \cite{delzanna_badnell:2014} rates are adopted here, together with the corresponding
A-values.  
The DW collision strengths and A-values for the extra 18 levels ($n$=6,7)
that were already present in CHIANTI, and were calculated by 
\cite{odwyer_etal:12_fe_9} have been retained for the present model ion.
The observed energies have been taken from the  \cite{delzanna:09_fe_8}
assessment and from NIST.


\subsection{Sc-like sequence}

\subsubsection{ \ion{Fe}{vi}}

We have included the $R$-matrix calculations of 
\cite{ballance_griffin:2008} for  \ion{Fe}{vi}. 
The target included 96  close-coupling levels arising from the 
3d$^3$,  3d$^2$ 4s, 3d$^2$ 4p, and 3p$^5$ 3d$^4$  configurations.
 They replace the previous calculations by 
\cite{chen_pradhan:1999} which included 80 close-coupling levels
from the 3d$^3$ and the 3d$^2$ 4s, 3d$^2$ 4p configurations.
The A-values for the allowed transitions are taken 
from \cite{ballance_griffin:2008}, while those for the 
forbidden transitions are retained from the previous CHIANTI model,
which originated from \cite{chen_pradhan:2000}.





\subsection{Cr-like sequence}

\subsubsection{\ion{Fe}{iii} }

\ion{Fe}{iii} is a new addition the the CHIANTI 
database. This is a relatively complex ion for which several 
 calculations and observations exist.
In terms of scattering calculations, we first considered the 
data of \cite{bautista_etal:2010} for the lowest 34 fine-structure levels
(mainly the 3d$^6$), and the Iron Project calculations of 
\cite{zhang:1996},  which included excitations to a selection of 
3d$^5$ 4s and 3d$^5$ 4p levels (for a total of 219 levels).
However, we have chosen to include the most recent large-scale
calculations of \cite{badnell_ballance:2014}, which included excitation to 
322 close-coupling levels, calculated with the Breit-Pauli $R$-matrix 
suite of codes. 
The authors did several comparisons with the results of other 
calculations, finding overall agreement, but suggested that the 
 Breit-Pauli calculations should be more accurate.

Experimental energies have been taken from NIST \citep{nist_v5} and 
from  \cite{ekberg:1993}. 
The ordering of the levels follows \cite{badnell_ballance:2014}.
The assignment of the experimental energies was straightforward for the 
lowest  levels (which produce the strongest lines for this ion),
however the  assignment to the higher levels was in several cases uncertain.
Some discrepancies between NIST and \cite{ekberg:1993} designations 
were also found.

We have also included, for consistency,  the corresponding A-values, which we 
recalculated with the same target adopted by \cite{badnell_ballance:2014} using 
{\sc autostructure} and the experimental energies whenever available.
We  included multipole orders up to three
(E2/M1+Breit-Pauli corrections and E3/M2).
Several previous calculations 
of radiative data also exist (cf. \cite{deb_hibbert:2009} for the 3d$^6$  levels,
and \cite{ekberg:1993} for 1908 transitions from the 
3d$^5$ 4s and 3d$^5$ 4p levels),  and some differences are found
for  the weaker transitions. 
Considering the forbidden transitions within the 
3d$^6$  levels, there is good agreement
between the  A-values calculated with {\sc autostructure} with those 
of \cite{deb_hibbert:2009}.


\section{Other improvements}

\subsection{DEM inversion software}

The CHIANTI\_DEM code has been modified substantially.
The input and output basic files are the same, as well as the 
format of how the line contribution functions can be stored.
The core inversion program has however been substantially modified.
The older $\chi^2$ minimization program has been replaced with a 
more robust  $\chi^2$ minimization program 
 written by M.Weber for the 
inversion of the Hinode XRT data (see, e.g. \citealt{weber_etal:2004}),
part of the Solarsoft program XRT\_DEM\_ITERATIVE2.

As an option, the routine can also 
run DATA2DEM\_REG, a routine written by I.Hannah
(described in \citealt{hannah_kontar:2012}) and/or the 
 Markov-Chain Monte-Carlo algorithm MCMC\_DEM,
written by V. Kayshap and described in \cite{kashyap_drake:98}.
In all three cases, the CHIANTI\_DEM code saves the inputs
as IDL save files, so users can subsequently rerun the inversion codes
independently of CHIANTI\_DEM. 
Details of the various parameters and where the programs can be 
downloaded can be found in the CHIANTI user guide and the 
CHIANTI\_DEM documentation on the CHIANTI website.

\subsection{Improvements to ChiantiPy}

\cite{landi_v7} provided the first published description of the ChiantiPy
software.  The goal of ChiantiPy is to provide the ability of access the
CHIANTI database and to compute line and continuum spectra.  ChiantiPy is
programmed in the Python programming language and is freely available from
https://www.python.org/.  The object oriented features of the language provide
a good match to the structure of the CHIANTI database which is organized around
single ions.  This is reflected  in ChiantiPy was the basic object is provided
by the {\em ion} class.

In addition to the {\em ion} class, two spectrum classes {\em spectrum} and
{\em mspectrum}, are available for calculation synthetic spectra for a group of
ions.  The {\em mspectrum} class implements the Python {\em multiprocessing}
module to allow multiple processors such as the available cores on a single cpu
or across a network.  Two new multi-ion class have been developed, {\em
ipymspectrum} and {\em bunch}.  The {\em ipymspectrum} is very similar to the
{\em mspectrum} except that it allows multiprocessing inside an IPython {\em
qtconsole} or a {\em noteboook}.  In the latter, the IPython session is
displayed inside a web browser.  IPython is freely available from
http://ipython.org.  The other new multi-ion class {\em bunch} simply
calculates the intensities, as well as other properties, for a selected set of
ions.

Several new methods have been developed, {\em intensityList()} and {\em
convolve()}.  The previous {\em ion} methods {\em intensityRatio()} and {\em
intensityRatioSave()} as well as the new method {\em intensityList()} are all
now inherited by the {\em ion} class and the multi-ion classes  {\em spectrum},
{\em mspectrum}, {\em ipymspectrum} and {\em bunch}.  Consequently, it is now
possible to calculate line intensity ratios as a function of temperature and
density for lines of different ions.

The multi-ion classes have been restructured so that one can specify to keep
all of the ion instances for later use.  For example, one can calculated the
spectrum selection of ions.  This spectrum can then be compared with the
spectrum of any of the ions in the initial selection.  The {\em convolve()}
methods can then perform a new convolution, with a different filter, of the ion
intensities available in the ion instances.  While useful, keeping all of the
ion instances can be memory intensive.  Consequently, the default option is not
to keep the ion instances.

Another improvement is the ability to specify the elemental abundance directly
to an {\em ion} instance and to specify the name of a file containing the
elemental abundances directly to any of the multi-ion instances.

The ChiantiPy software is freely available and can be downloaded from
http://sourceforge.net/projects/chiantipy.

\subsection{AIA responses}

\begin{figure}[!htbp]
\centerline{\epsfig{file=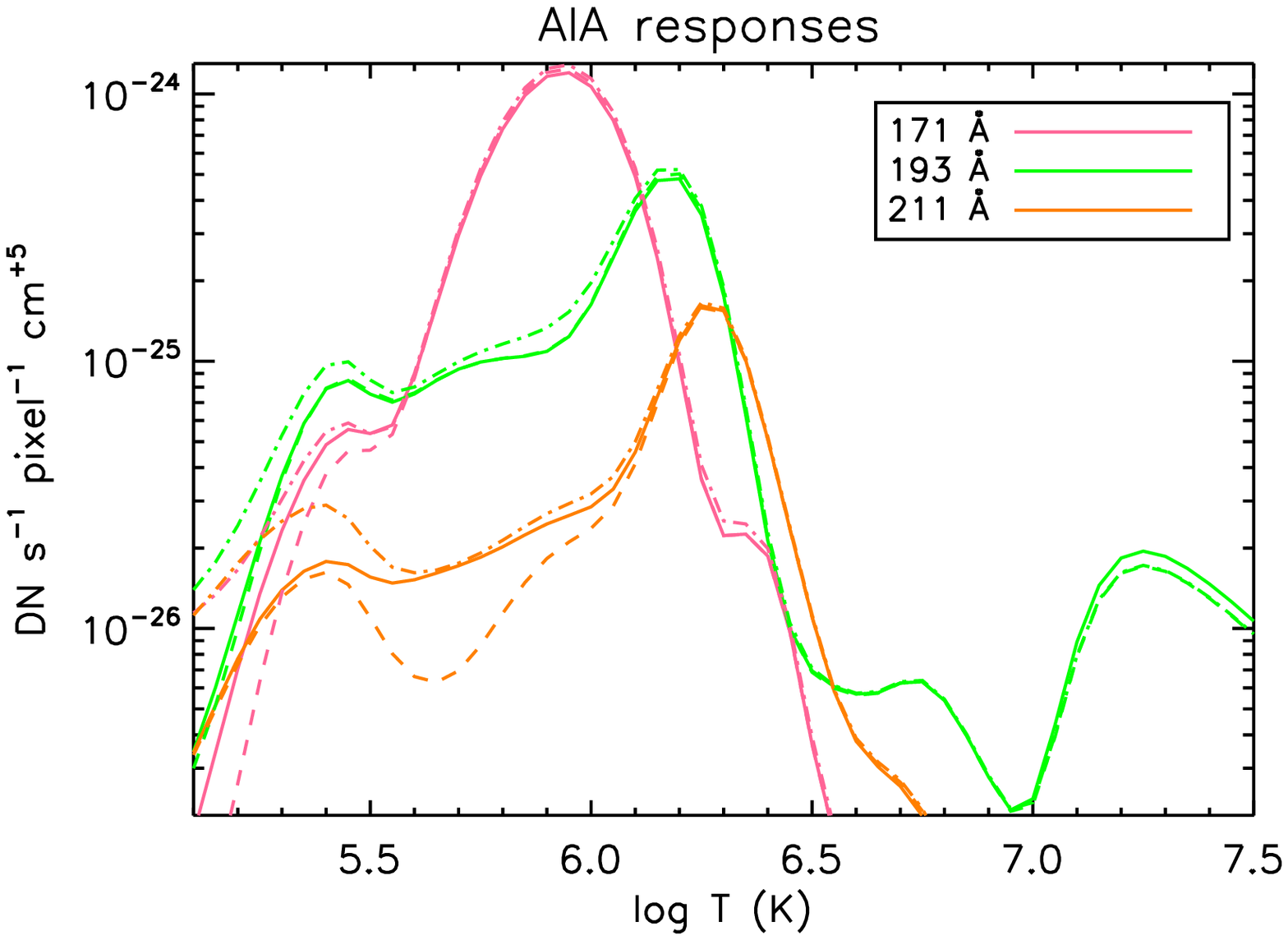, width=8.5cm,angle=0}}
\centerline{\epsfig{file=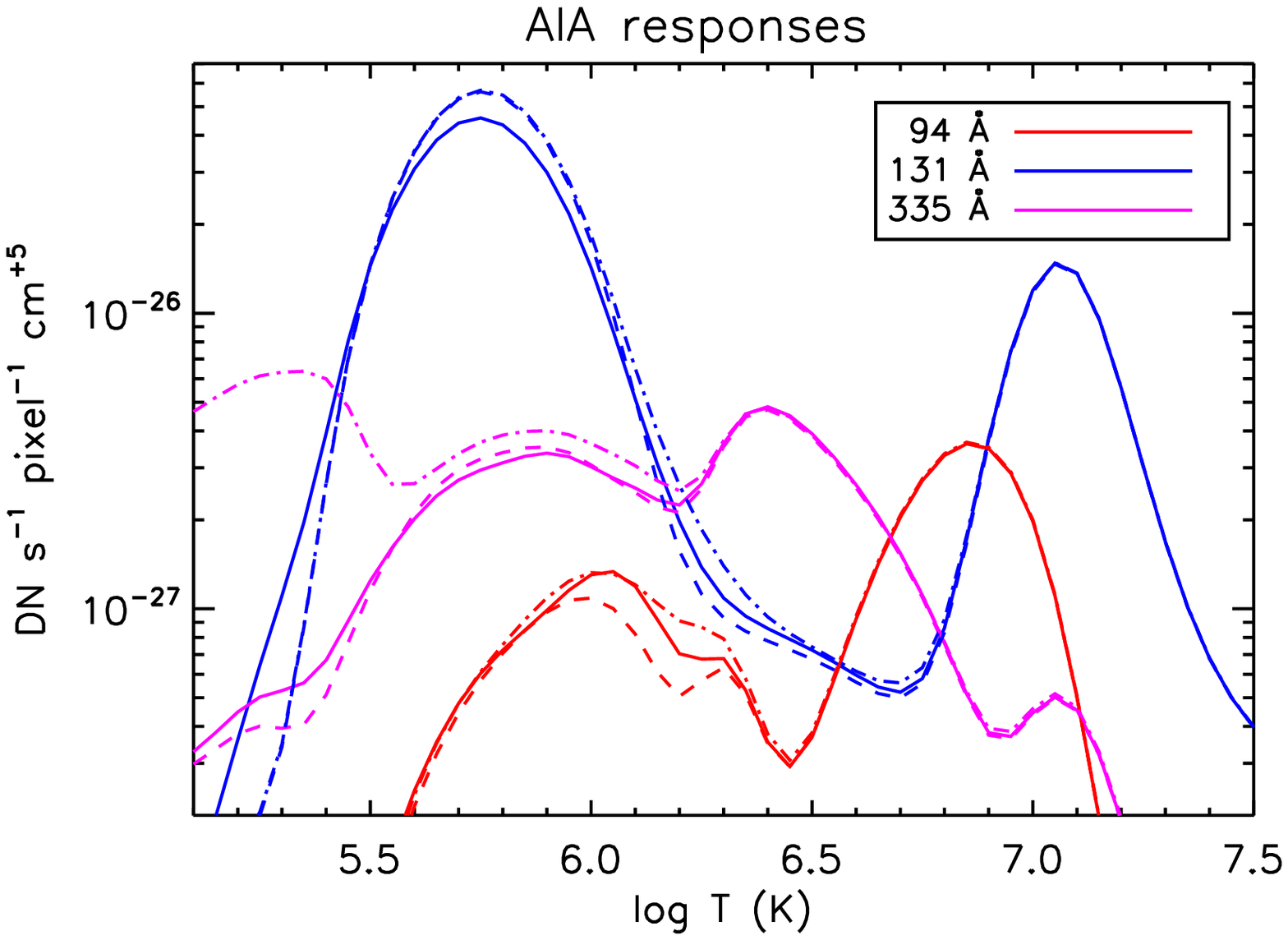, width=8.5cm,angle=0}}
\caption{SDO AIA  responses. Full line: present data; dashed line: CHIANTI v.7.1.4;
dot-dashed line: AIA software.}
\label{fig:aia_resp}
\end{figure}

Fig.~\ref{fig:aia_resp} shows the SDO AIA 
responses calculated with the present atomic data, 
a constant pressure of 10$^{15}$ cm$^{-3}$ K and 
a set of  `coronal' elemental abundances \citep{feldman:1992},
where the elements with low First Ionization Potential 
such as iron are increased by about a factor of four, 
compared to their photospheric abundances.

The new responses are shown together with 
those of the previous CHIANTI version 7.1.4, and those obtained from the 
AIA software, at the same 
pressure and with the same elemental abundances.  
We can see that the most significant changes occur for the 131 and 94~\AA\ bands,
for which the present data are a significant improvement.

Regarding the 131~\AA\ band, we note that for low-temperature 
plasma or the quiet Sun the dominant emission lines are the 
 \ion{Fe}{viii} 130.94 and 131.24~\AA\ transitions
\citep{odwyer_etal:10,delzanna_etal:11_aia}.
The previous version of CHIANTI had atomic data for this ion 
that were semi-empirically  adjusted by  \cite{delzanna:09_fe_8}.
We note, however, that the adjustment  was focused on improving 
the  collision strengths for the EUV  \ion{Fe}{viii} lines, and not those
populating the   3s$^2$ 3p$^6$ 4f $^2$F$_{5/2,7/2}$ levels, which 
give rise to the  130.94 and 131.24~\AA\ lines, respectively.

 The present atomic  data also represent an improvement 
for the SDO AIA 94~\AA\ band.
The previous version 7.1 of CHIANTI included  DW collision strengths 
for \ion{Fe}{x}. 
The \ion{Fe}{x} 94.012~\AA\ transition was underestimated by 
at least 30\%, as shown in 
\citep{delzanna_etal:12_fe_10}.
Also, as pointed out  in \cite{delzanna:2013_multithermal},
the intensity of the \ion{Fe}{x} 
 3s$^2$ 3p$^5$ $^2$P$_{3/2}$--3p$^6$ 3d $^2$D$_{5/2}$
line at 94.27~\AA\ was incorrect.
However, we recall that, as shown in \cite{delzanna:2013_multithermal}, 
a more  significant contribution to the band in active region cores
comes from an \ion{Fe}{xiv} 93.61\AA\ line, identified in 
 \cite{delzanna:12_sxr1}.
All the transitions observed within this band are now 
accounted for and have accurate atomic data, 
with the exception of a weak unidentified coronal line.

Finally, we caution users against  using AIA responses 
for the 94 and 131~\AA\ channel that are empirically adjusted, 
as done for example with the /chiantifix keyword, as described in 
\cite{boerner_etal:2014}. 
These empirical adjustments do not have any justification 
when one considers in detail the underlying atomic data.
For example, these adjustments modify the high-temperature peaks of the 
two bands, which are dominated by well-known lines from 
 \ion{Fe}{xviii} and  \ion{Fe}{xxi} \citep{odwyer_etal:10,petkaki_etal:12},
for which the atomic data are accurate.

\section{Conclusions}

The present version 8 of CHIANTI represents a significant step towards one of our goals,
to provide a consistent set of accurate atomic data for all the 
astrophysically important ions. 
Further work will focus on the ions in the Mg-like and Be-like 
sequences, for which effective collision strengths have recently been 
calculated with the  $R$-matrix ICFT method by the UK APAP network
\citep{apap-mg-like, apap-be-like}.
Inclusion of these data into CHIANTI has been delayed because 
\cite{aggarwal_keenan:2015} have recently strongly questioned the accuracy of the 
$R$-matrix ICFT approach for the Be-like ions. 
With a detailed comparison, \cite{be-like_rebuttal} have however
shown that the ICFT method produces the same results obtained 
with a fully relativistic Breit-Pauli $R$-matrix approach, which is 
needed for elements heavier than Zn. 
The results obtained by  Aggarwal and Keenan were actually more inaccurate
because of their smaller configuration-interaction / close-coupling 
expansion. 

We note that several other papers by Aggarwal and Keenan 
have also questioned the accuracy of the 
$R$-matrix ICFT approach for other ions,
but the UK APAP data included in the present version of 
CHIANTI have been selected following a careful assessment against 
other data-sets to ensure their accuracy.

The present version also represent a significant improvement 
for the important but very complex coronal iron ions, which produce 
a very large number of spectral lines across the electromagnetic spectrum,
and are used extensively for plasma diagnostic purposes.

The resonance structure near threshold often meant that 
 $R$-matrix collision strengths could not be fitted accurately
at low temperatures with a 5- or 9-point spline interpolation, so we have changed the 
the  format of the CHIANTI electron excitation files to 
store the data as they were calculated. 

The current calculations of the line intensities 
are still carried out considering each ion separately,
although ionization and recombination effects are taken into
account with some approximations.
The next version of CHIANTI will modify this approach.

\begin{acknowledgements}
A significant part of the present work was carried out by GDZ
 within SOLID,  to
 improve atomic data for spectral irradiance modelling.
SOLID (First European SOLar Irradiance Data Exploitation) is a collaborative 
SPACE Project under the Seventh Framework Programme (FP7/2007-2013) of the European Commission
under Grant Agreement N° 313188.  

GDZ and HEM acknowledge support  by the UK STFC bridging 
 grant of the DAMTP astrophysics group at the University of Cambridge.  \\

PRY acknowledges support from
NASA grant NNX15AF25G and NSF grant AGS-1159353.
The work of EL was supported by NASA grant NNX11AC20G
and NSF grant AGS-1154443.

The UK APAP network work was funded by STFC via 
the  University of Strathclyde (grant No.
 PP/E001254/1 and ST/J000892/1.)

We thank all the colleagues that have provided us 
with electronic material. In particular we thank 
Oleg Zatsarinny, 
Manuel Bautista, Cathy Ramsbottom, Nigel Badnell, 
Guyiun Liang, Martin O'Mullane.

We also thank Guyiun Liang for recalculating some of the atomic data
and  Nigel Badnell for advice on running {\sc autostructure}.

\end{acknowledgements}

\bibliographystyle{aa}

\bibliography{paper_rev}

\appendix

\section{New file formats}

\subsection{File format definitions for the ELVLC files}

The format for the elvlc file remained the same from CHIANTI 1 through to CHIANTI 7.1.
We have changed the format for version 8,  principally to allow level indices to go
beyond 999, and remove duplicate information. 
Only fine structure levels are considered in CHIANTI, and for each level the configuration, L, S
and J-values are given to describe the level. Energies are given in units of cm$^{-1}$ and both an
observed energy and a theoretical energy are given. If an observed energy does not exist, then the
CHIANTI software uses the theoretical energy for that level. 
If an observed energy is not available for a
level, then a “missing value” of -1 is assigned to that level.
The CHIANTI theoretical energies are not necessarily the same as those of the 
scattering calculations. In some cases, theoretical `best guess' energies are provided.
These are normally obtained by linear interpolation of the ab-initio level energies 
with the few experimental energies. The wavelength file has the corresponding wavelengths.
 Typical uncertainties of the ab-initio wavelengths are a 
 up to several \AA, while those obtained from  
the `best guess' energies are estimated to be around 1~\AA.

The energy file contains columns with a fixed number of characters, and the data entries are
terminated by a line containing only a “-1”. All subsequent lines are considered to be comments.

There are eight data columns and each is described below. The format for the column is indicated
by Fortran-style notation in  Table~A1. 
Additional columns can be present in the files. These are for information purposes and 
are not read by the CHIANTI software.

Levels can be arranged in any order, although following the observed or theoretical energy
ordering is recommended. Level 1 is recommended to be the ground level, although this is not
essential.
The usual  configuration format is, e.g., “3s2.3p2(2P).3d”. That is, orbitals are separated
by a “.” and parent terms are placed in brackets. However, separating orbitals by white space is
also present, e.g., “3s2 3p2(2P) 3d”.

We have added a level label string, which 
 can be used to attach a label to a level. 
An example is for Fe II for which the 
strings are used for 
multiplets in the same configuration which have the same LSJ labels.

\def\baselinestretch{1}
\begin{table}[!htbp]
\begin{center}
\caption{The format of the data in the energy files}
\begin{tabular}{llll}
\hline\hline\noalign{\smallskip}
Col. & Format & ID & Comment \\
\noalign{\smallskip}\hline\noalign{\smallskip}
1 & i7  & LVL &  Level index \\
2 & a30 & CONF  & Configuration description  \\
3 & a5  & LABEL  & Level label string \\
4 & i5  & 2S+1   & Spin multiplicity  \\
5 & a5  &  L     & Orbital angular momentum  \\
6 & f5.1&  J     & Total angular momentum   \\
7 & f15.3 & E\_o & Observed energy (cm$^{-1}$) \\
8 & f15.3 & E\_b & `Best-guess' theoretical energy (cm$^{-1}$) \\

\noalign{\smallskip}\hline
\end{tabular}
\end{center}
\label{tab:energies}
\end{table}

The elvlc file is read by the routine read\_elvlc.pro, which has
been modified to take into account the new format.
 In order to maintain
compatibility with the previous version of the routine,  read\_elvlc can be called
in the identical manner to the old   routine, i.e.,
\begin{verbatim}
IDL> read_elvlc, filename, l1,term, $
conf,ss,ll,jj,ecm,eryd,ecmth,erydth,ref
\end{verbatim}

However, read\_elvlc can also be called with:
\begin{verbatim}
IDL> read_elvlc, filename, elvlcstr=elvlstr
\end{verbatim}

where elvlcstr is an IDL structure containing the data.
Note that elvlcstr has the tag elvlcstr.data.energy which contains the 
best guess energy for a level.

\subsection{File format definitions for the SCUPS files}

The SCUPS files replace the SPLUPS files
in previous versions of the database. The format is very similar. 
For the new additions to version 8, 
this file contains the temperatures and  effective collision strengths,
scaled using the \cite{burgess_tully:92} [BT92] method.
The values as  originally calculated can be obtained by 
descaling the data in the files, using the IDL routine 
DESCALE\_ALL. 
In addition,  the collision strengths at scaled temperatures of 0 and 1 are given.
The value at threshold is extrapolated, while the value
at scaled temperature equal to 1 is either extrapolated or obtained from 
the high-energy limit.

For the other ions that have not been modified in version 8, 
the previous spline fits to the scaled
effective collision strengths are retained. Only the format of the files
is changed.

For each transition, there are three lines in the file.
The first line in the SCUPS files contains the information about the 
transition, the second the BT92-scaled temperatures, and the third
the BT92-scaled effective collision strengths,
as described in Table~A2. 
Note that even in the cases when a single temperature array 
was present, with the new format each transition will have a 
different scaled temperature array.

\def\baselinestretch{1}
\begin{table}[!htbp]
\begin{center}
\caption{The format of the data in the SCUPS files}
\begin{tabular}{llll}
\hline\hline\noalign{\smallskip}
Col. & Format & ID & Comment \\
\noalign{\smallskip}\hline\noalign{\smallskip}

1 & i7 & L1 & Lower level of transition (integer) \\
2 & i7 & L2 & Upper level of transition (integer) \\
3 & e12.3 & DE & Energy of transition, Rydberg (float) \\

4 & e12.3 & GF & Oscillator strength (float) \\
5 & e12.3 & LIM & High-temperature limit value (float) \\

6 & i5    & NT   & Number of scaled temperatures  \\

7 & i3    & T\_TYPE & BT92 Transition type  (integer) \\
8 & e12.3 & C\_VAL  & BT92 scaling parameter (float) \\

\noalign{\smallskip}\hline\noalign{\smallskip}

 & e12.3 & SCT  & Scaled temperatures (2nd line) \\

\noalign{\smallskip}\hline\noalign{\smallskip}
 & e12.3 & SCUPS  & Scaled effective collision strengths (3rd line) \\

\noalign{\smallskip}\hline
\end{tabular}
\end{center}
\label{tab:scups}
\end{table}

The comments in the file are at the end, bracketed by two lines 
containing only a “-1”.

The  SCUPS file is read into an IDL structure as follows:

\begin{verbatim}
IDL> read_scups, splfile, splstr
\end{verbatim}

where SPLSTR has two tags called INFO and DATA that are both structures.
The tags for SPLSTR.INFO are listed in Table~A3. 

\def\baselinestretch{1}
\begin{table}[!htbp]
\begin{center}
\caption{The  tags for SPLSTR.INFO}
\begin{tabular}{llll}
\hline\hline\noalign{\smallskip}
Tag & Data & Type \\
\noalign{\smallskip}\hline\noalign{\smallskip}

ION\_NAME & Ion  name (CHIANTI format) & String*1 \\
ION\_Z    & Atomic number & Integer*1 \\
ION\_N    & Spectroscopic number & Integer*1 \\
ION\_ROMAN & Ion name (Roman numerals) & String*1 \\
ION\_LATEX & Ion name (latex format) & String*1 \\
ION\_LATEX\_ALT & Ion name (alternative latex format) & String*1 \\
COMMENTS &  File comments  & String array \\
CHIANTI\_VER &  Version number & String*1 \\
TIME\_STAMP  &  Time file was made & String*1 \\
FILENAME & Filename (including path) & String*1 \\
MISSING  & Value for missing data & Float*1 \\
NTRANS   & Number of transitions  & Long*1 \\
\noalign{\smallskip}\hline
\end{tabular}
\end{center}
\label{tab:info}
\end{table}

The tags for SPLSTR.DATA are listed in Table~A4. 
The size of the TEMP and UPS arrays will
be set to  the maximum number of temperatures (NT\_MAX) in the dataset.
If e.g.  NT\_MAX=20 and a particular transition only has
upsilons defined for 10 temperatures, 
then TEMP[0:9] and UPS[0:9] will contain
these values, and TEMP[10:19] and UPS[10:19] will
be set to the ‘missing value’, defined in  UPSSTR.INFO.MISSING.

\def\baselinestretch{1}
\begin{table}[!htbp]
\begin{center}
\caption{The  tags for  SPLSTR.DATA }
\begin{tabular}{llll}
\hline\hline\noalign{\smallskip}
Tag & Data & Type \\
\noalign{\smallskip}\hline\noalign{\smallskip}

LVL1 & Lower level index & Integer*1 \\
LVL2 & Upper level index & Integer*1 \\
DE & Energy (Rydberg)  & Float*1 \\
GF &  Oscillator strength & Float*1 \\
LIM & High-temperature limit & Float*1 \\
T\_TYPE & BT92 Transition  & Integer*1 \\
C\_UPS & BT92  Scaling parameter & Float*1 \\
NSPL  & Number of temperatures & Integer*1 \\
STEMP & Scaled temperature values & Float*NT\_MAX \\
SPL &  Scaled effective collision strength values & Float*NT\_MAX \\
\noalign{\smallskip}\hline
\end{tabular}
\end{center}
\label{tab:sdata}
\end{table}

\end{document}